\documentclass[prl,twocolumn,aps,superscriptaddress,showpacs,letterpaper]{revtex4-1}
\usepackage{graphicx,psfrag}
\usepackage{epsfig}
\usepackage{natbib}

\begin{document}

\title{Universal thermopower of bad metals}

\author{V.~Zlati\'c}
\affiliation{Institute of Physics, Zagreb POB 304, Croatia}
\affiliation{Department of Physics, Georgetown University, Washington, DC 20057, USA}
\author{G.~R.~Boyd}
\author{J.~K.~Freericks}
\affiliation{Department of Physics, Georgetown University, Washington, DC 20057, USA}

\begin{abstract}
``Bad metals" have a large
linear resistivity at high-$T$ that is universally seen in oxides close to the Mott-Hubbard insulating phase. 
They also have an universal thermopower $\alpha(T)$: (i) at very low doping (lightly doped)  $\alpha(T)$ has a pronounced low-$T$ peak that shifts to higher-$T$ with doping; (ii) at moderate doping (underdoped) $\alpha(T)$ has a small low-$T$ peak that shifts to lower-$T$ with doping and has a high-$T$ sign change; and (iii) at the highest doping (overdoped) $\alpha(T)$ is negative and depends monotonically on $T$. Here we show that the simplified Hubbard model provides an easy to understand description of this phenomena due to the universal form for the chemical potential versus $T$ for doped Mott insulators and the applicability of the Kelvin formula for the thermopower.
\end{abstract}

\date{\today}
\maketitle

{\bf Introduction} ~
The high-temperature phases of complex oxides are attracting considerable current interest,  
as they exhibit a wide range of anomalous transport behaviors, typical of bad metals.  
Their resistivity is often (quasi) linear at high temperatures, continuing to increase as a function 
of temperature beyond the minimal metallic conductivity of the Mott-Ioffe-Regel limit~\cite{hussey_2004}   
and corresponding to a mean free path on the order of one unit cell, or less. 
The  thermopower of cobalt oxides~\cite{terasaki.1997,lee_2006,Kaurav_2009}, 
vandadates~\cite{urano.2000}, and ruthenates~\cite{okamoto.2008}  is often much 
larger than seen in normal metals, making them interesting for high-temperature thermoelectric 
applications~\cite{smart_materials_2008}, like recovering waste heat.
The cuprates, often thought of as paradigmatic bad metals, have an unusual thermoelectric response
which correlates with the superconducting properties~\cite{cooper_1992,honma_2008,cooper.2011}. 
Their thermopower $\alpha (T)$ exhibits universal features~\cite{cooper_1992,honma_2008,cooper.2011} 
and is extremely sensitive to doping. 
Its room temperature value, $\alpha (T_{RT})$ is a good measure of the number of holes $\delta$. 
In the underdoped regime,  $\alpha (T_{RT})$ depends exponentially on $\delta$, 
at optimal doping (highest $T_c$), $\alpha (T_{RT})$ changes sign, 
and in the overdoped regime, $\alpha (T_{RT})$ is a linear function of $\delta$. 

The defining feature of these bad metals  is a large but metallic resistivity 
and a thermopower which grows to much larger values than in normal metals. 
Since the mean free path inferred from the Drude formula drops below one lattice 
spacing~\cite{hussey_2004}, it is difficult to describe charge and heat transport in 
terms of quasiparticle currents, and  the Fermi liquid paradigm fails.  
In the phase diagram of bad metals, one often finds a nearby insulating region 
with a Mott-Hubbard gap in the excitation spectrum, indicating the importance of the on-site correlations.   
Considering the enormous difference in the structural properties of various cuprates 
used in the above studies, the universal behavior of the thermopower and linear resistivity 
suggest that the transport 
properties of  the high-temperature phase might be described  by an effective  band of strongly 
correlated electrons (similar to the spin-1/2 Anderson model describing the Kondo effect in real materials). 
The solution of that model then provides the appropriate conceptual framework for more realistic materials modeling.

Here, the transport properties of bad metals are explained by the exact solution 
of the simplified Hubbard model (spinless Falicov-Kimball model). 
For large correlation, we find that the density of states (DOS) has a Mott-Hubbard gap 
and the optical conductivity is characterized by the transfer of spectral weight from the low-energy 
Drude peak to the high-energy incoherent background.  The resistivity is linear and the thermopower
has anomolous behavior.
This is precisely what is seen in the Hubbard model at high temperature 
and it is the universal features of this bad metal phase that we describe here.
Several recent papers used a single band model to discuss the effects of strong correlation 
on the transport properties  of bad metals~\cite{mckenzie_2000,zlatic_2012,Shastry_PhysRevB.87.035126}. 
The advantage of the simplified Hubbard model is that it admits an exact solution that is numerically tractable at any doping.

\begin{figure}
\includegraphics[width=0.95\linewidth,clip]{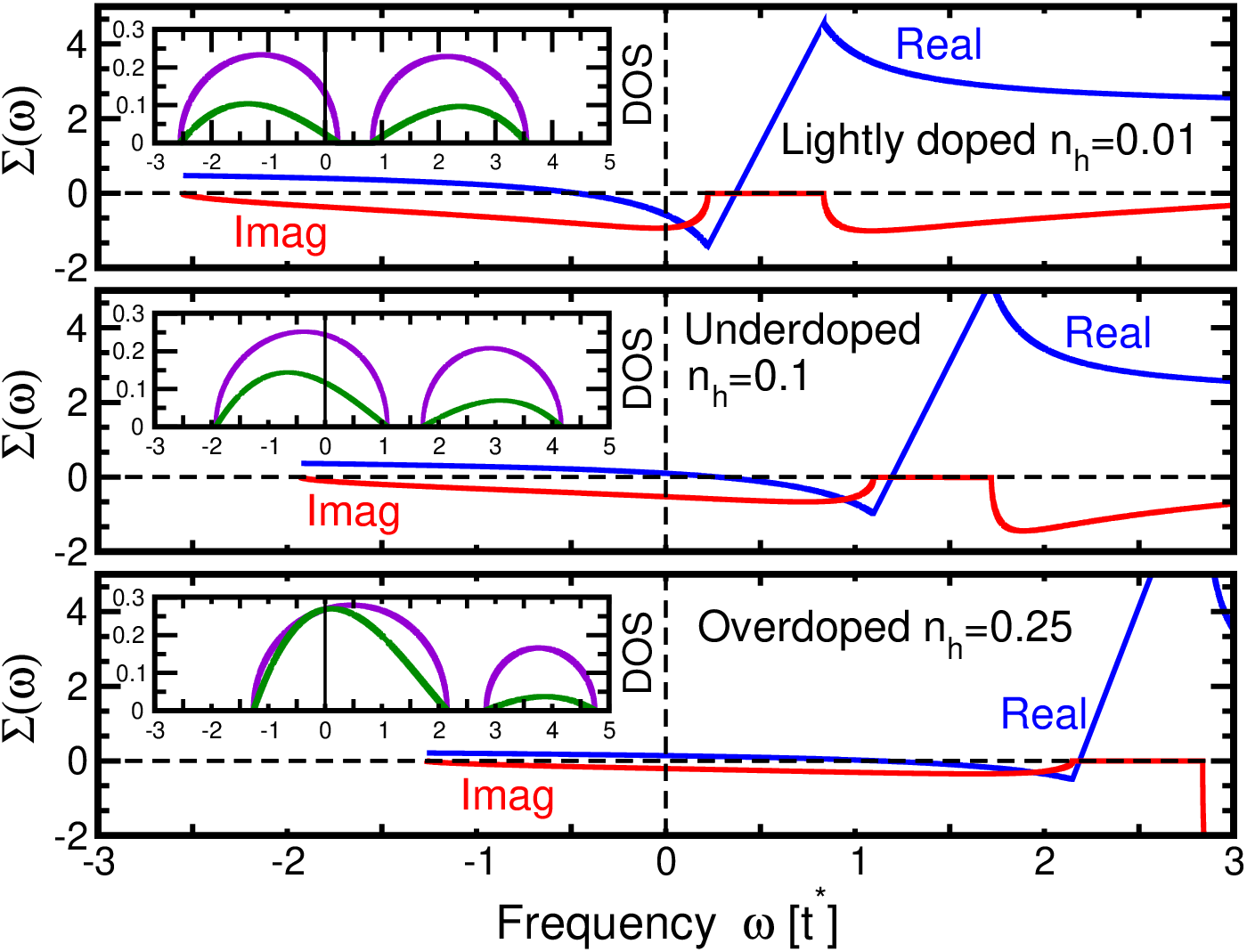}\\
\includegraphics[width=0.85\linewidth,clip]{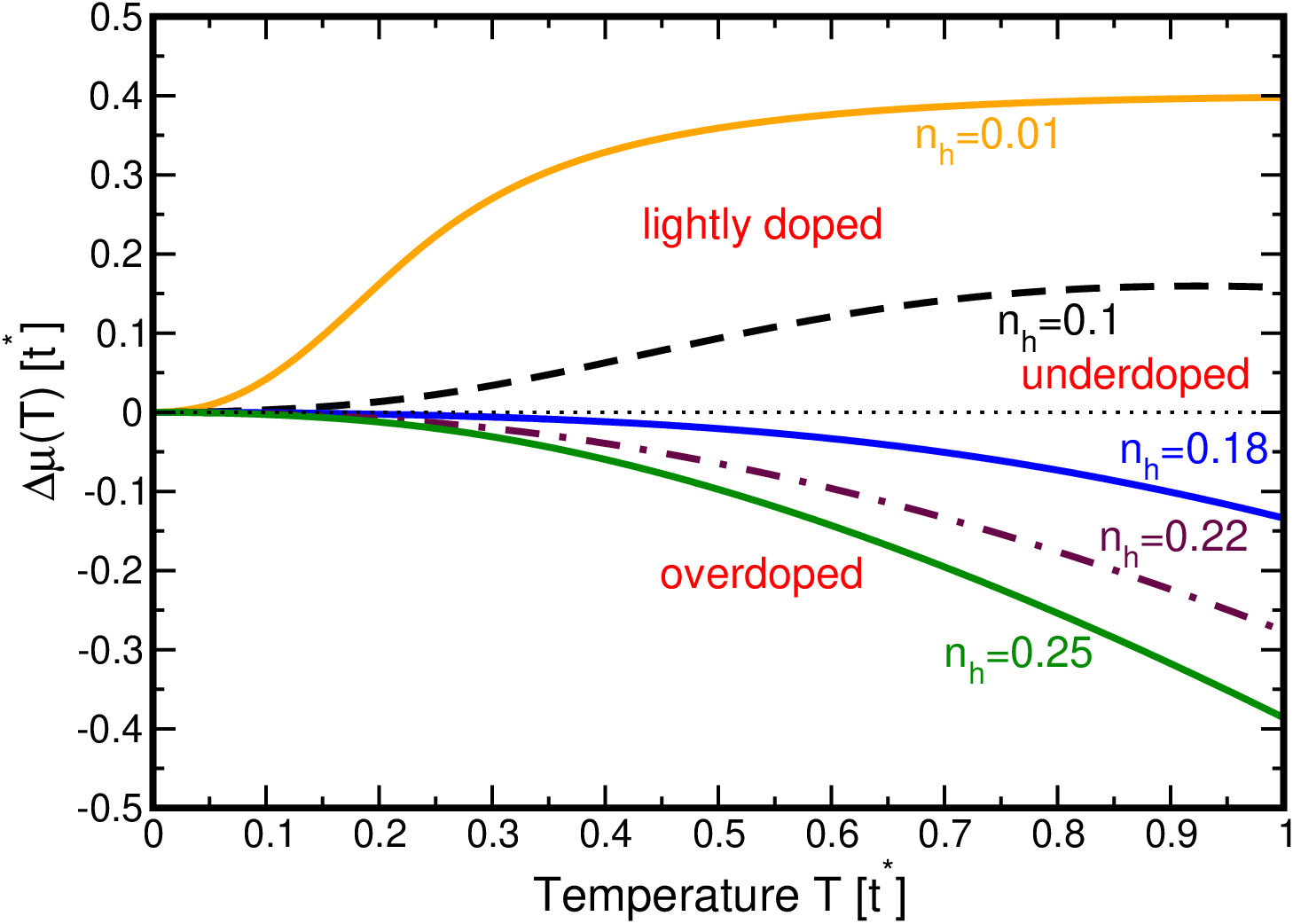}
\caption[]{(color online) 
Upper panels: 
the real part (blue) and the imaginary part (red) of the self energy are plotted versus energy 
for $U=3$ and for various dopings. The energy is measured with respect to the zero-temperature chemical potential $\mu_0$, in units of $t^*$. 
The inset shows the local DOS (violet) and the transport DOS (green) at $T=0$. 
Lower panel:  
the chemical potential, measured with respect to its zero-temperature value, 
plotted versus temperature for various dopings.
}
\label{Bethe_self_energyFig}
\end{figure}

{\bf Model and calculations}~
The model is defined by the Hamiltonian 
\begin{equation}
H=-\frac{t^*}{\sqrt Z}\sum_{ij}c_i^{\dagger}c_j + U \sum_i w_i c_i^{\dagger} c_i 
\end{equation}
where the summation is over $N$ lattice sites, 
$c_i^{\dagger} \;( c_i)$ is the itinerant electron creation (annihilation) operator and $w_i=f^\dagger_if^{}_i$ is 1 or 0
and represents the number operator of a localized electron on site $i$. $U$ is the interaction strength and $t^*$ is the
hopping scaled so that we can properly take the infinite coordination number ($Z\rightarrow \infty$)  
limit (the hopping is between nearest-neighbors only). 
We work on a Bethe lattice, measure the energy with respect to the zero-temperature chemical potential, $\mu_0$, 
and rescale the Hamiltonian by  $t^*$, which makes the Green's functions and the self energy dimensionless.
Localized electrons are distributed according to an annealed thermodynamic ensemble and 
$\sum_i w_i/N=w_1$ is their average filling. 
There are $N_c$ itinerant electrons per site and we take $N_c=w_1$ to describe the simplified Hubbard model in a paramagnetic phase. 
We dope away from half-filling, so $w_1= 0.5-\delta = N_c$, where  $\delta$ is the doping (or density of holes $n_h$).
At high enough temperatures, the results obtained for $w_1= N_c$ 
provide a good approximation to the Hubbard model (by taking the up spin electrons as the mobile 
electrons and the down spin electrons as the localized ones). 
The two models differ at low-$T$ where coherence sets into the Hubbard model 
creating Fermi-liquid phases or different forms of ordered phases. 

The solution is obtained by employing dynamical mean field theory 
(DMFT)~\cite{Vollhardt,DMFTRMP,RevModPhys.75.1333} 
in  the infinite coordination limit $Z \rightarrow \infty$. We focus on the local retarded
Green's function $G_{\rm loc}(\omega)$, defined in the standard way, and utilize the conventional DMFT algorithm
to formulate the solution~\cite{ RevModPhys.75.1333}.
On the  Bethe lattice, $G_{\rm loc}(\omega)$ satisfies a cubic equation~\cite{vanDongen_1997} that we solve numerically 
and the self energy is then given by the expression  
\begin{equation}
\Sigma(\omega)= \omega+\mu- G_{\rm loc}(\omega) -{1\over  G_{\rm loc}(\omega) }
~
\label{Sigma}
\end{equation}
where we set $t^*=1$. 
\begin{figure*}[thb!]
\centering
\includegraphics[width=0.80\columnwidth,clip]{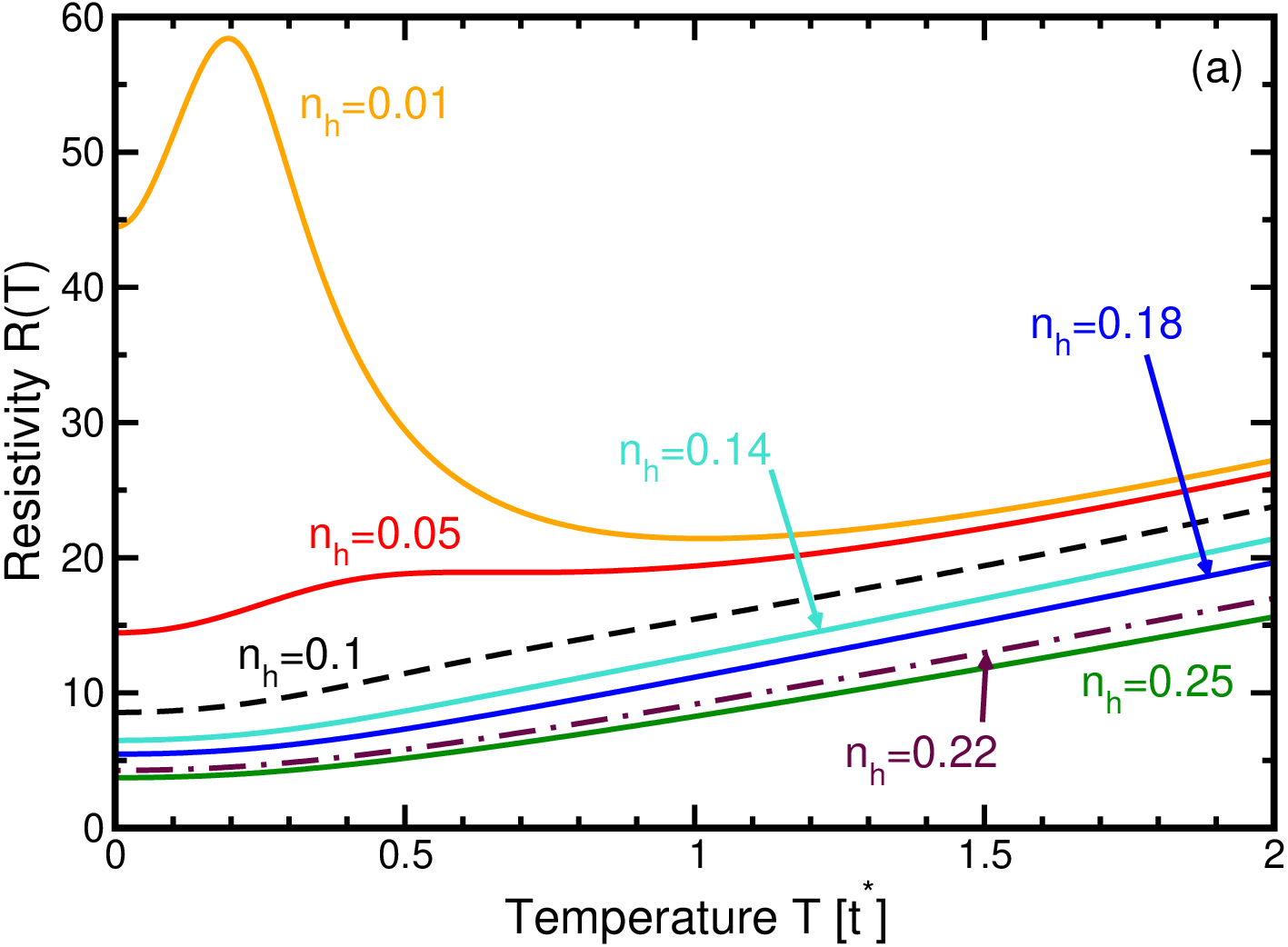} ~~
\includegraphics[width=0.80\columnwidth,clip]{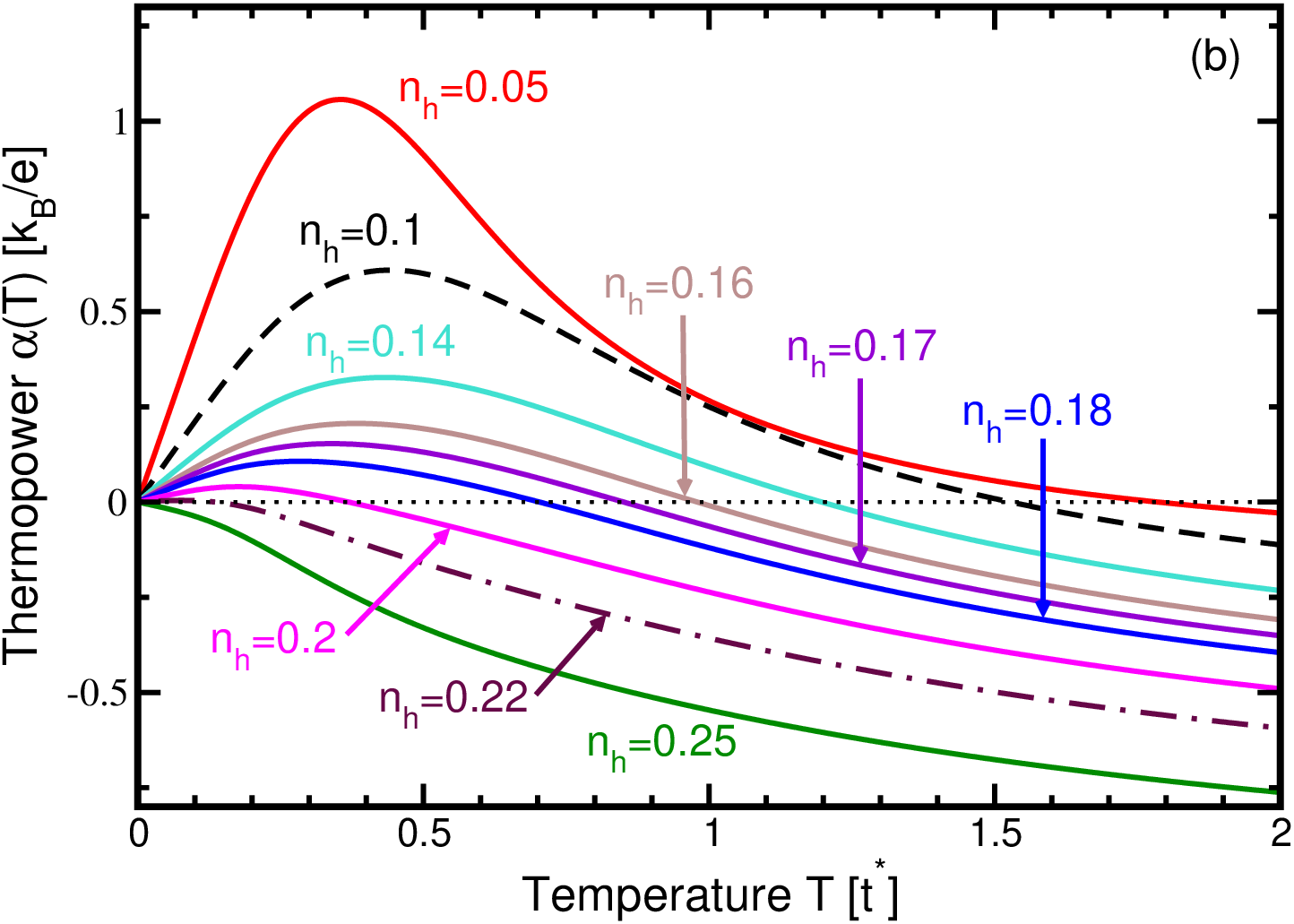}
\\
~~
\\
\includegraphics[width=0.80\columnwidth,clip]{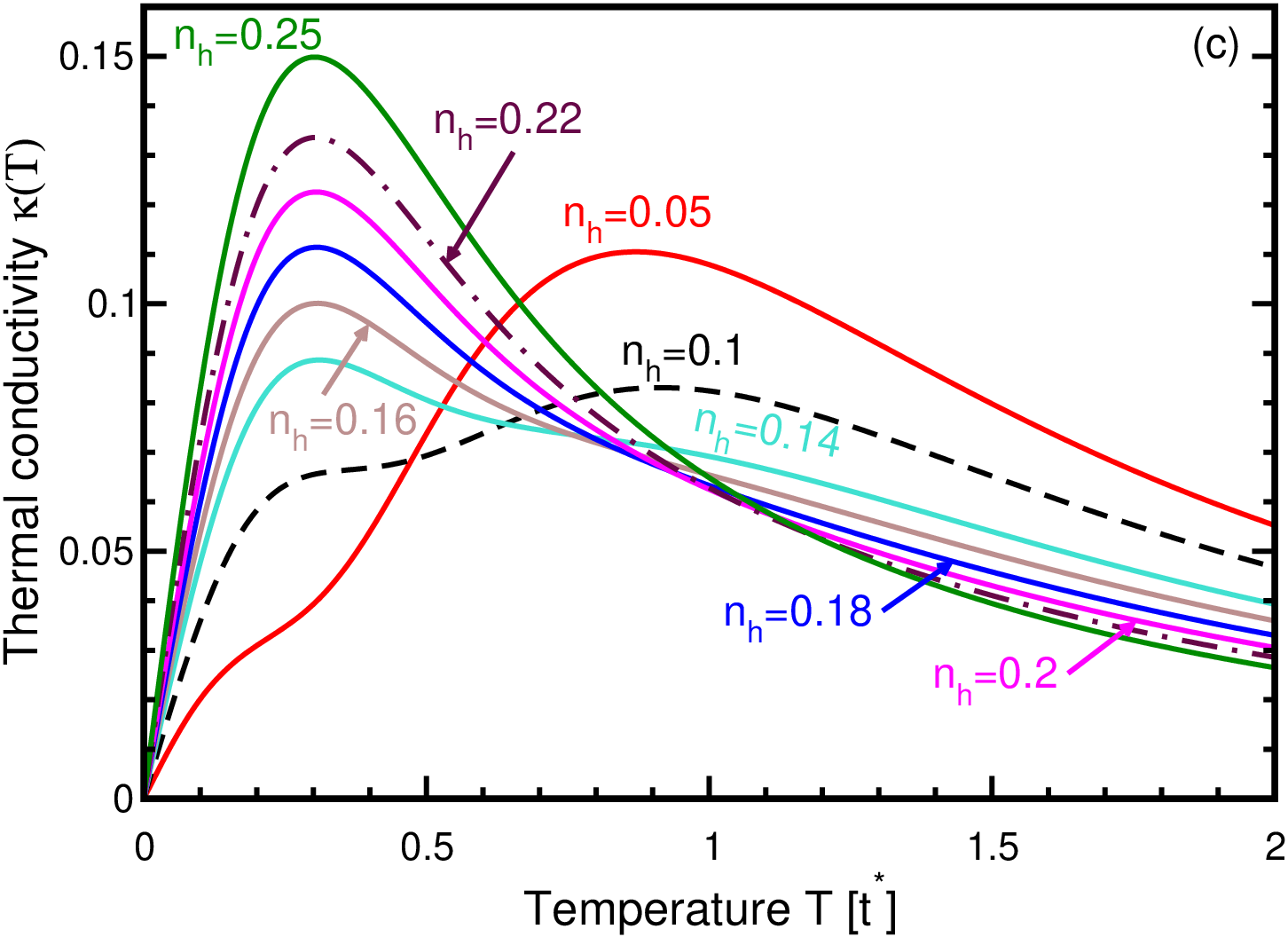} ~~
\includegraphics[width=0.80\columnwidth,clip]{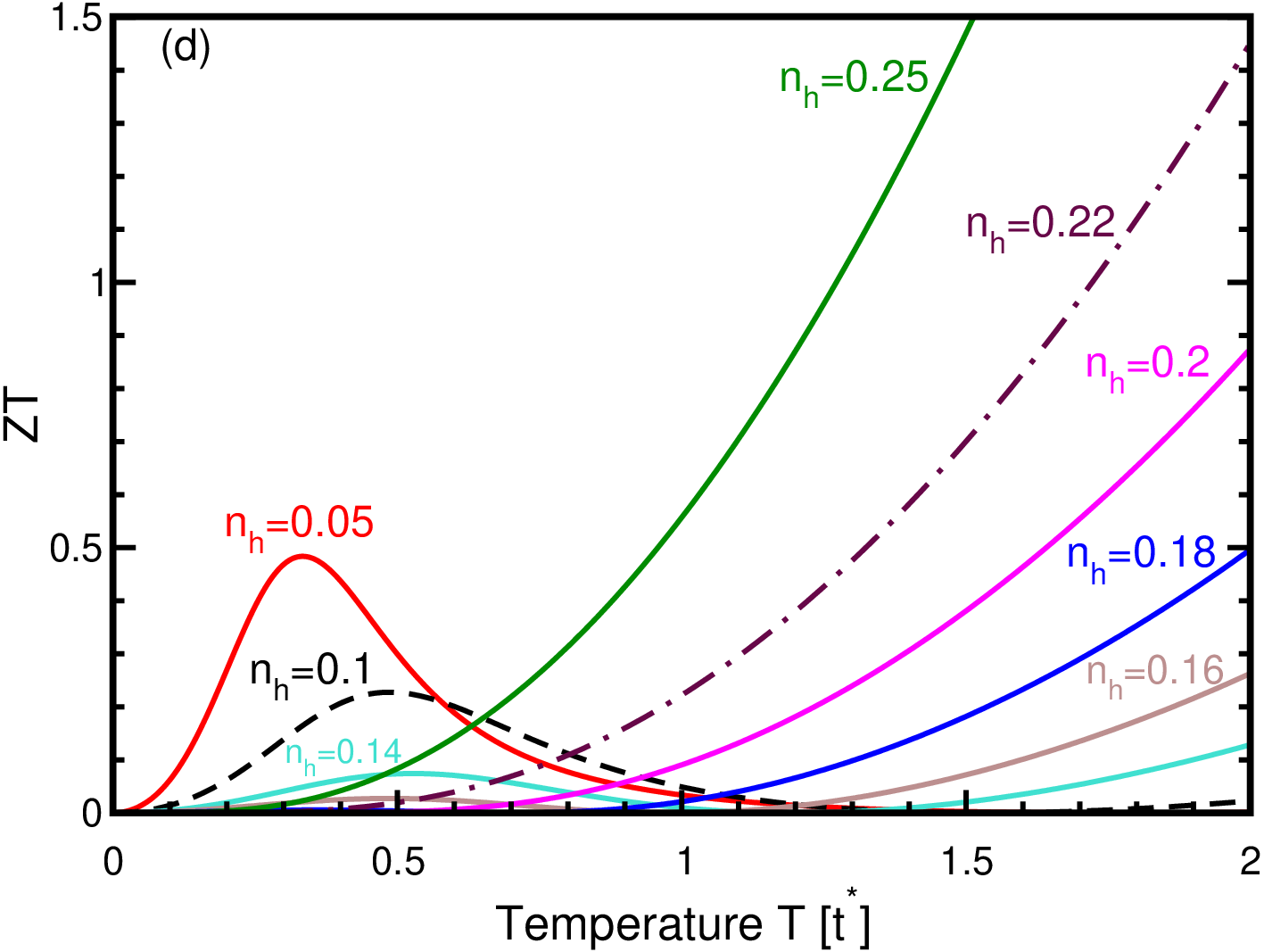}
\caption[]{(color online) (a) The resistivity $R(t)$, (b) thermopower $\alpha(T)$, (c) thermal conductivity $\kappa(T)$ and (d) figure-of-merit $ZT$
plotted versus temperature for $U=3$ and for various dopings, as indicated in the panels.
The boundary between doped insulators and bad metals is indicated by the dashed line;  
the boundary between the underdoped and overdoped regions is indicated by the dashed-dotted line.} 
\label{Bethe_transportFig}
\end{figure*}

The electrical conductivity $\sigma$, the  thermopower $\alpha$,  
and the thermal conductivity $\kappa_e$ are calculated by linear response theory~\cite{RevModPhys.75.1333}.
This gives
$\sigma(T)= \sigma_0 L_{11}(T)$, 
$\alpha(T)=\left({k_B}/{e}\right) {L_{12}}/{T L_{11} }$, and 
$\kappa_e=\left({k_B}/{e}\right)^2 (\sigma/T)\left( L_{22}/L_{11}-L_{12}^2/L_{11}^2 \right)$, 
where $\sigma_0=e^2/(\hbar Z a)$ with $a$ an effective lattice spacing and 
\begin{eqnarray}
L_{mn}(T)       
=
\int_{-\infty}^{\infty}d\omega~
\left(-\frac{\partial f(\omega)}{\partial \omega}\right)\ \omega^{m+n-2} \ 
\Lambda^{}_{tr}(\omega)~. 
                           \label{eq: transport_integrals}                             
\end{eqnarray} 
The transport function is 
\begin{eqnarray}
\Lambda^{}_{tr}(\omega)
=
\frac{4}{3\pi^2}
\int d\epsilon ~
\Phi(\epsilon)
[\rm Im ~ G^{}_{}(\epsilon,\omega)]^2~, 
                      \label{eq: lambda_tr}
\end{eqnarray}
where $\epsilon$ is the noninteracting band energy, 
$\Phi(\epsilon)$ is the noninteracting transport DOS (DOS weighted by the square of the velocity),  
and $G^{}_{}(\epsilon,\omega)$ is the dimensionless Green's function of conduction electrons calculated within DMFT.
Using 
$\Phi(\epsilon)= (4-\epsilon^2)\sqrt{4-\epsilon^2}/{2\pi}$, 
 the transport function can be calculated exactly, with the result ~\cite{joura}
\begin{equation}
\Lambda_{\rm tr}(\omega)=\frac{1}{3\pi^2}\rm{Im}^2[G_{loc}(\omega)]\left(
\frac{|G_{loc}(\omega)|^2-3}{|G_{loc}(\omega)|^2-1} \right) 
~.
\end{equation}

The numerical data show that the thermopower is well approximated by the Kelvin expression~\cite{shastry_2010}, 
$\alpha_K(T)= -(k_B/q_e) (\partial\mu/\partial T)$.  
This expression is obtained by assuming that the diffusion of $\delta n$ particles of 
charge $q_e$, due to the temperature difference $\Delta T$, gives rise to an entropy change $\delta s$ 
and a voltage difference $\Delta V$.  
In a stationary state with zero current, the loss of the thermal energy, $\delta s\times \Delta T$, is balanced 
by the work done by the electrical field, $q_e\delta n\times \Delta V$,  such that 
$\Delta V /\Delta T= (1/q_e) (\delta s/ \delta n)$ and the Kelvin expression for $\alpha_K(T)$ then follows from 
the appropriate Maxwell relation. Unlike the Kubo formula, which is derived by making the driving fields uniform 
before they become static, the Kelvin formula is based on equilibrium thermodynamics and the static limit is taken first.  
As pointed out by Arsenault et al.~\cite{Shastry_PhysRevB.87.035126}, 
a good agreement between $\alpha_K(T)$ and $\alpha(T)$ can be taken as an indication 
that transport properties are mainly determined by the equilibrium  fluctuations, 
i.e., by the renormalized DOS.  
The effects coming from the velocity factors or relaxation time are then of secondary importance. 

{\bf Results} ~
The numerical calculations are performed for $U=3$ which produces a Mott-Hubbard insulator at half filling.  
In the upper panel of Fig.~\ref{Bethe_self_energyFig}, we show, the frequency dependence 
of the zero-temperature self energy, the local density of states, $\rho(\omega)=-\mathrm{Im}~G_{\rm loc}(\omega)/\pi$,  
and the transport function $\Lambda_{tr}(\omega)$, for various concentrations of holes.
At very low doping, the slopes of $\mathrm{Re}~\Sigma(\omega)$ and $\mathrm{Im}~\Sigma(\omega)$ 
at $\omega=0$ (i.e., at the chemical potential) are very large, such that the quasiparticle cannot be defined. 
As $\delta$ increases, $\mathrm{Re}~\Sigma(0)$ decreases, while $\mathrm{Im}~\Sigma(0)$ 
increases rapidly up to a maximum. 
For $\delta > \delta_s$, both $\mathrm{Re}~\Sigma(0)$  and $\mathrm{Im}~\Sigma(0)$ decrease 
with $\delta$ and the low-energy part of $\mathrm{Im}~\Sigma(\omega)$ becomes a linear function of $\omega$.  
For large $\delta$, $\mathrm{Im}~\Sigma(0)$ becomes very small and an approximate description 
in terms of (dirty) quasiparticles becomes possible~\cite{PhysRevLett.110.086401}. 

The temperature dependence of the chemical potential,  
obtained from the condition $N_c=\int d\omega f(\omega) \rho(\omega)=0.5-\delta$, 
is shown in the lower panel of Fig.~\ref{Bethe_self_energyFig}. 
For small doping, the low-temperature values of $\mu(T)$ are just below the band edge 
of the lower Hubbard band. An increase of temperature shifts $\mu(T)$ across the band edge, 
towards the center of the gap, which is typical of a doped Mott insulator~\cite{zlatic_2012}. 
At higher doping, $\mu_0$ is closer to the center of the Hubbard band and 
$\mu(T)$ grows slowly towards a maximum. However, for large enough doping, $\delta > \delta_s$, 
$\mu(T)$ never crosses the band edge and the model describes an underdoped bad metal
($\delta_s\simeq 0.1$ for $U=3$).
A further increase of doping reduces the initial slope and the high-temperature maximum of $\mu(T)$, 
until they both vanish at the critical doping, $\delta=\delta_c$, which separates the `underdoped' and 
the `overdoped' regimes ($\delta_c\simeq 0.22$ for $U=3$). 
At $\delta_c$, $\mu(T)$ is nearly constant over an extended temperature range 
and the entropy, considered as a function of doping, assumes a local maximum;  
the maximum of $\Lambda_{tr}(\omega)$ is now close to $\omega=0$ and  the thermopower is negligibly small. 
In the overdoped regime,  $\delta > \delta_c$, there is a further shift of  $\mu_0$  away from the center 
of the lower Hubbard band  and $\mu(T)$ decreases monotonically, as in a Fermi liquid with the same density of holes. 

The local DOS and the transport function are shown in the insets of the upper panel 
of Fig.~\ref{Bethe_self_energyFig}. Doping increases the weight of the lower Hubbard band 
with respect to the upper one, and shifts the maximum of $\rho(\omega)$ away from the 
maximum of $\Lambda_{tr}(\omega)$.  For constant $\delta$, the  local DOS and the transport function 
of the Bethe lattice are temperature-independent, except for a shift given by $\mu(T)$. 

The  temperature dependence of $R(T)$, $\alpha(T)$, $\kappa(T)$ and 
the electronic figure of merit  $ZT$ at various dopings is shown in Fig.~\ref{Bethe_transportFig}. 
The transport functions exhibit three different behaviors, depending on the level of doping.
For $\delta\leq \delta_s(U)$, 
the  low-temperature transport is not affected by the gap but at intermediate temperatures, 
when $\mu(T)$ crosses the band edge, the asymmetry  of the electron and hole states is much enhanced. 
This gives rise to large maxima of $R(T)$ and $\alpha(T)$,  the 
break-down of the Wiedemann-Franz law, and a large $ZT$. 
The signature of lightly doped insulators are the pronounced peaks in $R(T)$ and $\alpha(T)$, and the shifts 
of these peaks to higher temperatures for higher doping~\cite{zlatic_2012}. 

In  the underdoped region, $\delta\geq \delta_s$,  transport is completely determined 
by the incoherent excitations in the lower Hubbard band. 
The peak in $R(T)$ is suppressed and the linear resistivity extends to very low temperatures. 
The thermopower has a low-temperature peak but its height decreases rapidly with $\delta$. 
Unlike in the lightly doped Mott insulators, the peak of $\alpha(T)$ in (underdoped) bad metals shifts with $\delta$ 
to lower temperatures and, at high temperatures, $\alpha(T)$ changes sign. 
The low-temperature peak of $ZT$ is rapidly suppressed with doping but at higher temperatures $ZT$ 
becomes large.  

In the overdoped region, $\delta\geq \delta_c$, the resistivity is further reduced  
but the onset of the linear region is pushed to higher temperatures. Below the linear 
region, $R(T)$ exhibits $T^2$ behavior.   
The initial slope of $\alpha(T)$ is now negative and $\alpha(T)$, like $R(T)$, is a  monotonic function 
of temperature. $ZT$ is very small at low temperature but it grows to large values at high temperatures. 
Here, Im $\Sigma$ is sufficiently small that the transport properties of bad metals 
can be described in terms of  `resilient quasiparticles' \cite{PhysRevLett.110.086401} or by a dirty Fermi liquid.  

\begin{figure}
\includegraphics[width=0.85\columnwidth,clip]{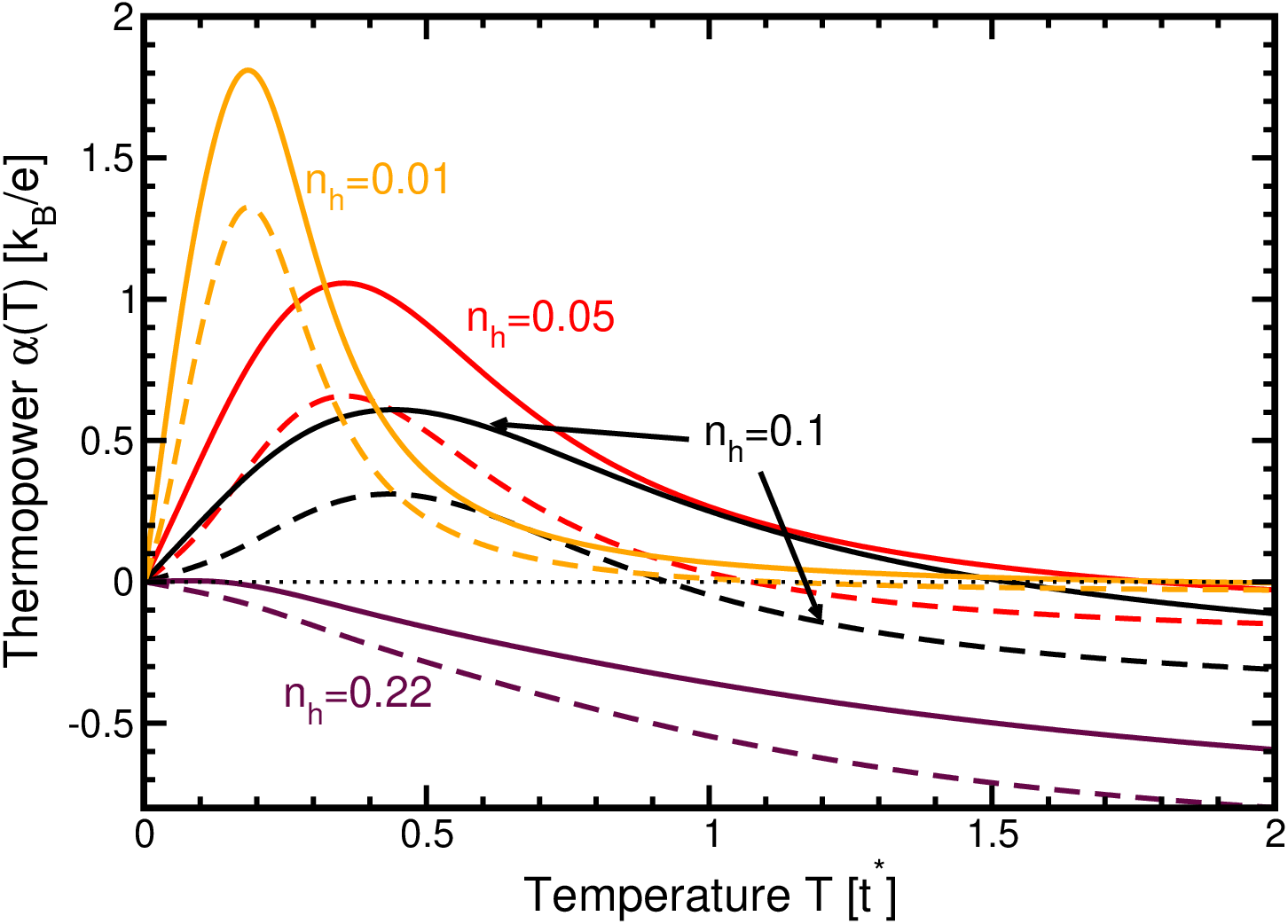}
\caption[]{(color online) The thermopower obtained from the Kubo formula (full line) 
is compared with the Kelvin formula (dashed line) for various dopings.} 
\label{kelvinfig}
\end{figure}

The comparison between the thermopower calculated 
by the Kelvin and Kubo formula is shown in Fig.~\ref{kelvinfig}. 
The semiquantitative agreement between $\alpha(T)\simeq \alpha_K(T)$ indicates that the thermodynamic fluctuations 
are the main cause of transport anomalies and that the steady-state thermal transport 
is directly related to $\mu(T)$. This is at the heart of the universal features seen in the thermopower, 
since the chemical potential is a function of the DOS and doped Mott insulators at high temperature 
share similar DOS for a wide range of different models. It does not require quasiparticles or `resilient quasiparticles'.

{\bf Summary}~ 
We studied the transport properties of bad metals at various dopings using the DMFT  
solution of the simplified Hubbard model.  Since the self-energy functional of this model 
is known exactly, we found the transport properties at arbitrary doping and obtained 
the difference between overdoped bad metals, described by resilient quasiparticles, 
and underdoped bad metals, where the quasiparticle concept breaks down.
We also studied a slightly doped Mott-Hubbard insulator, which is currently not numerically possible 
for the systems described by the Hubbard or the Anderson model.  
In general,  we find a linear resistivity and a large thermopower, as observed in many bad metals 
mentioned in the introduction. These features are the result of large fluctuations induced by the on-site 
Coulomb repulsion, which also gives rise to the transfer of spectral weight in the optical conductivity,  
described elsewhere, together with the other dynamical properties of the model. 

We find that the simplified one band Hubbard model describes the temperature and 
doping dependence of the thermoelectric response of cuprates  surprisingly well. Taking $T=0.05$ as room temperature 
(assuming the bandwidth $D=4=2$ eV), we obtain the following features: 
(i) the values of  $\alpha(T)$ at $T_{RT}=$290 K increase exponentially, 
when $\delta$ is reduced much below $\delta_c$; 
(ii)  $\alpha(T_{RT})$ changes sign at $\delta_c$;  and
(iii)  for $\delta > \delta_c$,  $\alpha(T_{RT})$ 
becomes a linear function of $\delta$. All these features are seen in experiments 
on cuprates~\cite{cooper_1992,honma_2008,cooper.2011} which also show that 
$\delta_c$ gives the maximum superconducting $T_c$. 
The universal nature of the behavior is tied to the similar behavior of $\mu(T)$ 
for different doped Mott insulators and the fact that the thermopower is 
determined primarily by the Kelvin formula in this parameter regime.

\section{Acknowledgements}

JKF and VZ were supported by the National Science Foundation under grant number DMR-1006605 for the data analysis and interpretation. JKF and GRB were supported by the Department of Energy, Office of Basic Energy Sciences, under grant number DE-FG02-08ER46542 for the development of the codes and numerical analysis. The collaboration was supported the Department of Energy, Office of Basic Energy Sciences, Computational Materials and Chemical Sciences Network grant number DE-SC0007091. JKF was also supported by the McDevitt bequest at Georgetown University.

\bibliography{biblio}

\end{document}